\documentclass{edm_article}

\usepackage{xcolor}
\usepackage{tikz}
\usepackage{tabularx}
\usepackage{pgfplots,pgfplotstable}
\usepackage{comment}
\newcommand\blfootnote[1]{%
  \begingroup
  \renewcommand\thefootnote{}\footnote{#1}%
  \addtocounter{footnote}{-1}%
  \endgroup
}

\begin{document}
\title{Prescribing Deep Attentive Score Prediction Attracts Improved Student Engagement}

% \author{
% Anonymous\\
%       \affaddr{Anonymous Institution}\\
%       \email{anonymous@anonymous.edu}
% }
\author{
Youngnam Lee\textsuperscript{*1},
Byungsoo Kim\textsuperscript{*1},
Dongmin Shin\textsuperscript{1},
JungHoon Kim\textsuperscript{1},\\
Jineon Baek\textsuperscript{1,2},
Jinhwan Lee\textsuperscript{1}
Youngduck Choi\textsuperscript{1,3}
\\
\affaddr{
    \textsuperscript{1}Riiid! AI Research, \textsuperscript{2}University of Michigan, \textsuperscript{3}Yale University
}\\
\email{
\{yn.lee, byungsoo.kim, dm.shin, junghoon.kim, jineon.baek, jh.lee, youngduck.choi\}@riiid.co
}
}

\maketitle

\begin{abstract}
Intelligent Tutoring Systems (ITSs) have been developed to provide students with personalized learning experiences by adaptively generating learning paths optimized for each individual.
Within the vast scope of ITS, score prediction stands out as an area of study that enables students to construct individually realistic goals based on their current position.
Via the expected score provided by the ITS, a student can instantaneously compare one's expected score to one's actual score, which directly corresponds to the reliability that the ITS can instill.
In other words, refining the precision of predicted scores strictly correlates to the level of confidence that a student may have with an ITS, which will evidently ensue improved student engagement.
However, previous studies have solely concentrated on improving the performance of a prediction model, largely lacking focus on the benefits generated by its practical application.
In this paper, we demonstrate that the accuracy of the score prediction model deployed in a real-world setting significantly impacts user engagement by providing empirical evidence.
To that end, we apply a state-of-the-art deep attentive neural network-based score prediction model to \emph{Santa}, a multi-platform English ITS with approximately 780K users in South Korea that exclusively focuses on the TOEIC (Test of English for International Communications) standardized examinations. 
We run a controlled A/B test on the ITS with two models, respectively based on collaborative filtering and deep attentive neural networks, to verify whether the more accurate model engenders any student engagement.
The results conclude that the attentive model not only induces high student morale (e.g. higher diagnostic test completion ratio, number of questions answered, etc.) but also encourages active engagement (e.g. higher purchase rate, improved total profit, etc.) on \emph{Santa}.
\end{abstract}


\blfootnote{*Equal contribution.}

\section{Introduction}
The significance that standardized examinations (e.g. SAT and TOEIC) currently hold is to provide an objective criteria in which each individual's academic performance is measured.
Accordingly, Intelligent Tutoring Systems (ITSs), which generate optimized learning paths for each student, often include functions such as estimating expected performance on standardized examinations.
In this regard, measuring the expected academic performance of a student has become an interesting area of study in Artificial Intelligence in Education (AIEd). 
These studies focus on modelling a student's understanding of a target subject based on their learning activities.
For instance, Matrix Factorization (MF) \cite{iqbal2017machine,polyzou2016grade,morsy2017cumulative,morsy2019sparse,rechkoski2018evaluation,hu2018course,ren2018ale,ren2019grade} is a prevalent method used for grade prediction, in which the latent vectors of students and courses are learned by factorizing a student-grade matrix into two low-rank matrices. 
Markov and semi-Markov models are also some other popular approaches for grade prediction \cite{jo2018time,hu2018course,rechkoski2018evaluation}.
With the advances in deep learning, neural network based models with deeper hidden layers, such as Multi-Layer Perceptron, Recurrent Neural Networks and Convolutional Neural Networks, were introduced to predict student's academic performance \cite{patil2017effective,hu2019reliable,jo2018time,hu2019academic}.
In \cite{choi2020assessment}, the Transformer-based \cite{vaswani_2017} bidirectional encoder model was first pre-trained to predict masked assessments and then fine-tuned to predict exam score, resulting in a state-of-the-art score prediction model.
Although precision of academic performance prediction is significant as it is directly associated to a reliability of an ITS, previous studies have mainly focused on improving the accuracy of the prediction, leaving discussion about the benefits of precise prediction on student engagement fairly opaque.

In this paper, we direct our attention towards the correlation between the precision of score prediction and student engagement.
Our study starts by hypothesizing that students will show higher level of engagement if they experience a more precise score prediction while interacting with ITS.
We empirically verify our hypothesis on \emph{Santa}, a multi-platform English ITS with approximately 780K users in South Korea that exclusively focuses on the TOEIC (Test of English for International Communications) Listening and Reading Test Preparation.
In the experimental studies, we run a controlled A/B test with two score prediction models that differ in accuracy, which are respectively based on collaborative filtering with Mean Absolute Error (MAE) of 78.9 and deep attentive neural networks with MAE of 49.8.
The results show that the superior performing, deep attentive neural network based score prediction model induces more student engagement. 
These benefits range from ones that are derived from learning behavior (e.g. preliminary test completion ratio, membership rates, the average number of questions a student answered after the diagnostic test) to more active engagement (e.g. purchase rate, average revenue per user, and total profit).
To the best of our knowledge, this is the first work studying the benefits of accurate score prediction of ITS on student engagement.

\begin{figure*}[t]
\begin{center}
\includegraphics[width=0.9\textwidth]{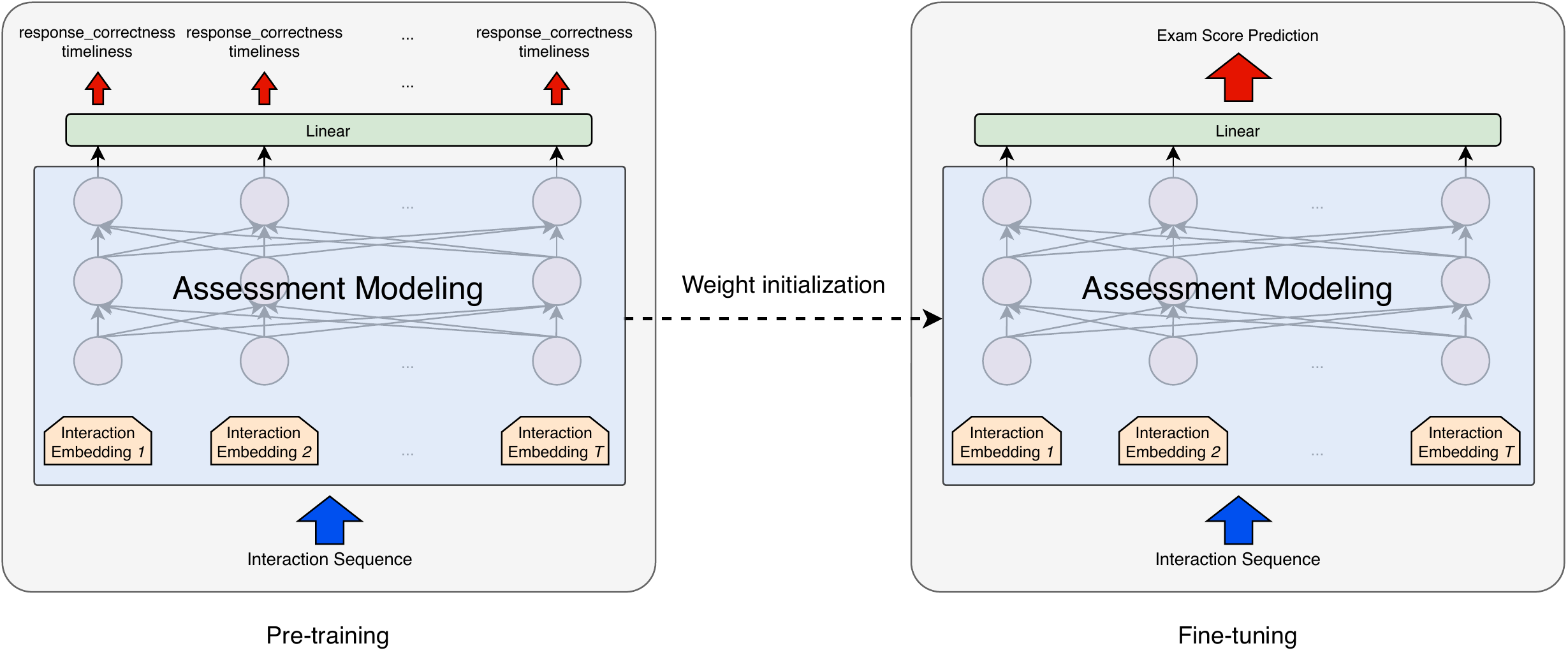}
\end{center}
\caption{Pre-training/fine-tuning scheme of Assessment Modeling for score prediction.
First, a model is pre-trained to predict two assessments: response correctness and timeliness.
After pre-training, the last layer of the model is replaced with a layer with randomly initialized weights and appropriate dimension for score prediction. The parameters in the model are fine-tuned to predict exam scores.}
\label{fig:model_arc}
\end{figure*}

\section{Related Works}
The related works of this study can be grouped into two categories: academic performance prediction and student engagement.

\subsection{Academic Performance Prediction}
Predicting a student's academic performance is a significant aspect in solving the problems within AIEd. A successful prediction model can be used to recommend appropriate courses, provide interventions for at-risk students, and optimally allocate learning materials. Extensive work has been conducted on performance prediction, exploring a wide range of methodologies from simple regressions to deep learning. 

The most widely used methodology in grade prediction is low rank Matrix Factorization (MF) \cite{iqbal2017machine,polyzou2016grade,morsy2017cumulative,morsy2019sparse,rechkoski2018evaluation,hu2018course,ren2018ale,ren2019grade}. Low rank MF assumes that there is a low-dimensional latent space containing features that can effectively represent both students and the academic tasks students will be graded on. These features can be interpreted as representations of a student's knowledge. We find these features by decomposing a student-grade matrix into a product of two low-rank matrices. The authors of \cite{polyzou2016grade} show that the MF-based model outperforms other course/student-specific regression models.
\cite{morsy2017cumulative} improved the model by assuming that different courses share a common latent feature space, since the totality of a student's knowledge should not change based on the courses they are taking. 

Markov and semi-Markov models are another popular set of models for grade prediction \cite{jo2018time,hu2018course,rechkoski2018evaluation}.
These models capture the dynamic evolution of a student's learning status and leverage it to effectively predict outcomes.
\cite{hu2018course} develops course-specific hidden Markov and semi-Markov models for grade prediction.
\cite{jo2018time} models student behavior in MOOCs by using Hidden Markov Models (HMMs) and Multinomial Mixture Models (MMMs) to cluster sequences of student actions. 
The study applies an LSTM model to predict the students' final grades.
Markov models are also used to estimate a student's performance on educational games \cite{tadayon2019predicting} or to predict student retention in MOOCs \cite{balakrishnan2013predicting}.

\cite{patil2017effective,hu2019reliable,jo2018time,hu2019academic} introduce deep-learning based prediction models.
The authors of \cite{hu2019reliable} introduce two types of Bayesian deep learning models for grade prediction using Multi-Layer Perceptron and LSTM architectures.
Their results show that their model outperforms several baseline models (including MF-based models and course-specific regression models) in detecting at-risk students. 
The authors of \cite{choi2020assessment} propose Assessment Modeling (AM), a pre-training method applicable to general ITSs.
In AM, a model is first pre-trained to predict several assessments of a student automatically made by ITS during one's learning process.
Their results show that a Transformer \cite{vaswani_2017} based neural network model with AM improves model accuracy compared to the same network with other state-of-the-art pre-training methods (such as BERT \cite{devlin2018bert} based word embedding and QuesNet \cite{yin2019quesnet} question embedding) on exam score prediction and review correctness prediction. 

\subsection{Student Engagement}
% analyzing engagement
Student engagement is also an actively studied topic in the field of AIEd.
Several works have analyzed student engagement patterns to figure out which factors vastly impact engagement. 
\cite{warner2015high} studied how people use digital textbooks and compare engagement patterns among high school students, college students, and online website viewers.
\cite{mulqueeny2015improving} investigated student engagement in an online learning system which outperformed a traditional classroom on key indicators of engagement, such as time on-task, engaged concentration, and boredom.
\cite{slater2016semantic} found correlations between semantic features of mathematics problems and indicators of engagement.
\cite{li2016understanding} discriminated behavioral engagement and cognitive engagement, and argued that most of students who were behaviorally engaged were not cognitively engaged.

% predicting engagement level
Another line of student engagement research focused on predicting engagement level.
\cite{okur2017behavioral} proposed a two-phased approach for automatic engagement detection, which utilized contextual logs and appearance information to infer behavioral engagement.
\cite{naik2019analyzing} investigated the relationship between engagement and performance.
Firstly, this work analyzed log traces for each learner to calculate engagement indicators that represent learner's engagement level.
Based on the quantified engagement indicators, prediction on the learner's performance were attempted.

% enrollment
Enrollment is a sign of strong engagement since it involves determination that a student must invest to take a certain course.
Accordingly, predicting and promoting enrollment is highly relevant to student engagement research.
\cite{gonzalez2018inferring} proposed a novel extension of Factorization Machines to infer students' course enrollment information from incomplete data.
\cite{bydvzovska2016course} presented a course enrollment recommender system which recommended selective and optional courses based on students' skills, knowledge and interests.
\cite{slim2018predicting} identified factors that affect the likelihood of enrolling.
This work analyzed the enrollment predictability of such factors using logistic regression, support vector machines, and semi-supervised probability methods.

% mooc
With the development of Massive Open Online Courses (MO OCs), several works studied student engagement in a MOOCs environment.
\cite{labarthe2016does} proposed a recommender which provides each student with an individual list of contacts based on their own profile and activities to foster their engagement in MOOCs.
\cite{lui2017exploring} investigated the relationship between students' self-evaluation of their previous knowledge and students' engagement behaviors in MOOCs through a polytomous item response theory model.

\section{Score Prediction Models}
Our studies are based on comparing the two approaches for score prediction: a collaborative filtering based approach and Assessment Modeling.
The following subsections briefly cover each approach.
More detailed descriptions can be found in \cite{lee2016machine} and \cite{choi2020assessment}.

\subsection{Collaborative Filtering based Approach}
There are two phases in the Collaborative Filtering (CF)-based score prediction approach.
First, the CF-based model developed in \cite{lee2016machine} estimates the probability that a student responds correctly to each potential question.
In this model, each user or question is represented as a $k$-dimensional latent vector, where $k$ is the number of hidden concepts. 
For instance, if there are $n$ users with $m$ questions, we have user vectors $L_1, L_2, \cdots, L_n$ and question vectors $R_1, R_2, \cdots, R_m$ each with dimension $k$.
The knowledge level of user $i$ understanding question $j$ is represented as $X_{ij} = L_i \cdot R_j$.
Accordingly, the probability of user $i$ getting question $j$ correct is modeled as
$$\phi(X_{ij}) = \phi_a + \frac{1 - \phi_a}{1 + e^{-\phi_c (X_{ij} - \phi_b)}},$$
where $\phi_a$, $\phi_b$, and $\phi_c$ are parameters appropriately set, independently of questions or users.
The learning algorithm then finds the maximum likelihood estimator by minimizing the negative of log-likelihood of observed user-question entries with Frobeinus norm regularizer terms through the projected stochastic gradient descent.

Given the response correctness probabilities calculated from the CF-based model, scores for Listening Comprehension (LC) and Reading Comprehension (RC) are calculated through the following quadratic equations
\begin{align*}
score_{LC} = \theta_2x_{LC}^2 + \theta_1x_{LC} + \theta_0 \\
score_{RC} = \theta_5x_{RC}^2 + \theta_4x_{RC} + \theta_3,
\end{align*}
where $x_{LC}$ and $x_{RC}$ are each the average of predicted response correctness probability of potential questions in LC and RC, and $\theta$s are properly set parameters.
The final score is the sum of $score_{LC}$ and $score_{RC}$.

\subsection{Assessment Modeling}
\cite{choi2020assessment} introduced Assessment Modeling (AM), a fundamental pre-training method for general class of ITSs.
The motivation behind the works of AM is to deal with label-scarce problems in AIEd.
Score prediction is a typical example of such label-scarce educational problems since standardized exam scores are not obtainable within ITS.
Collecting the exam scores involves student action taken outside ITS.
The approach proposed in \cite{choi2020assessment} is based on a pre-training/fine-tuning paradigm.
In the pre-training phase, the Transformer-based \cite{vaswani_2017} bidirectional encoder model is trained to predict randomly masked assessments, which are interactive educational features available in ITS.
Examples of these assessments include response correctness (whether a student provides a correct response to a given question) and timeliness (Whether a student responds to each question within the time limit specified by domain experts).
In the fine-tuning phase, the last layer of the pre-trained model is replaced with a randomly initialized layer with an appropriate dimension for a specific downstream task.
Afterwards, the parameters in the model are updated to predict labels in the downstream task.
In the experimental studies conducted on EdNet \cite{choi2019ednet}, AM outperformed pre-training methods that learn the contents of learning materials in several downstream tasks including score prediction.
See Figure \ref{fig:model_arc} for graphical description of AM.

\begin{figure*}[h]
\begin{center}
\includegraphics[width=0.8\textwidth]{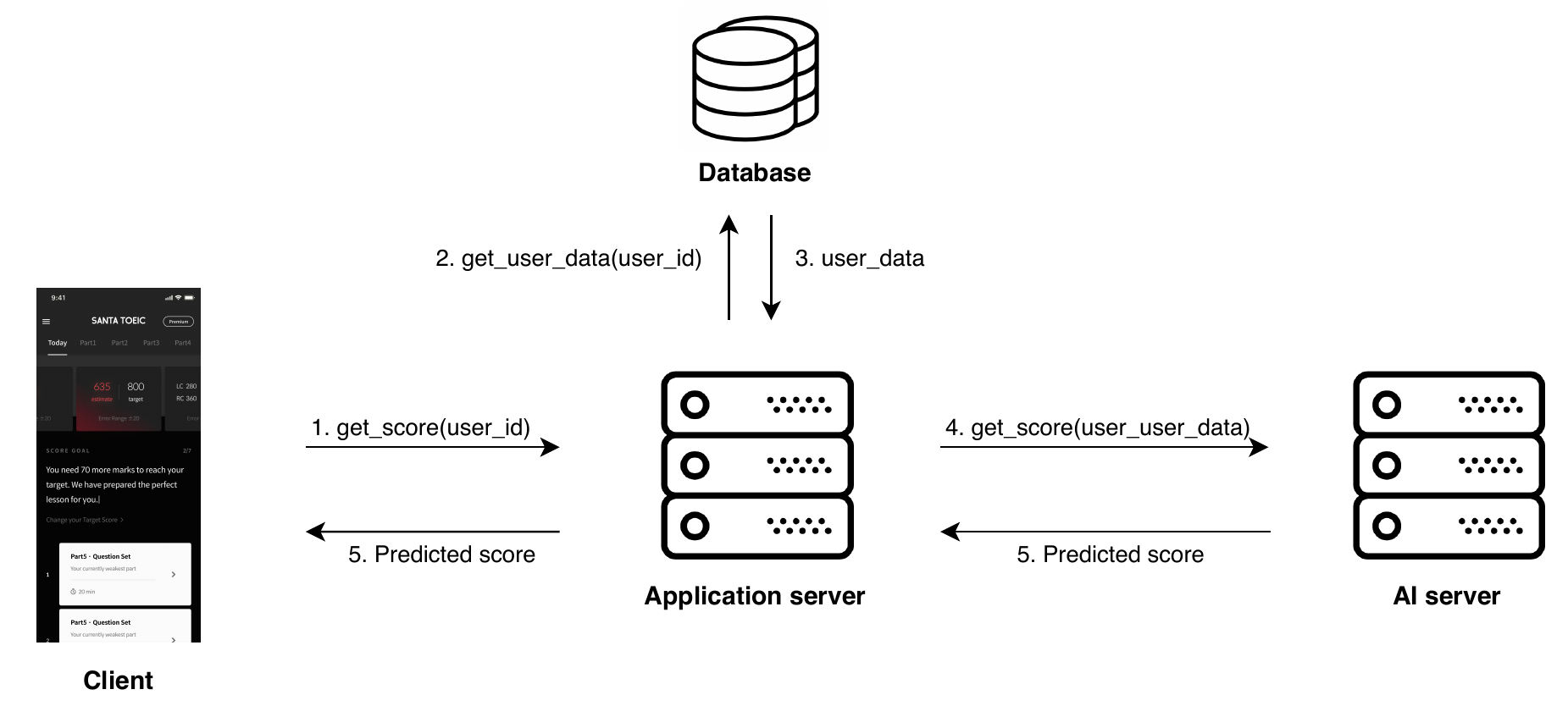}
\end{center}
\caption{The flow of score prediction.}
\label{fig:model_flow}
\end{figure*}

\section{experiments}
\subsection{Santa service}
\begin{figure}[!b]
\begin{center}
\includegraphics[width=0.45\textwidth]{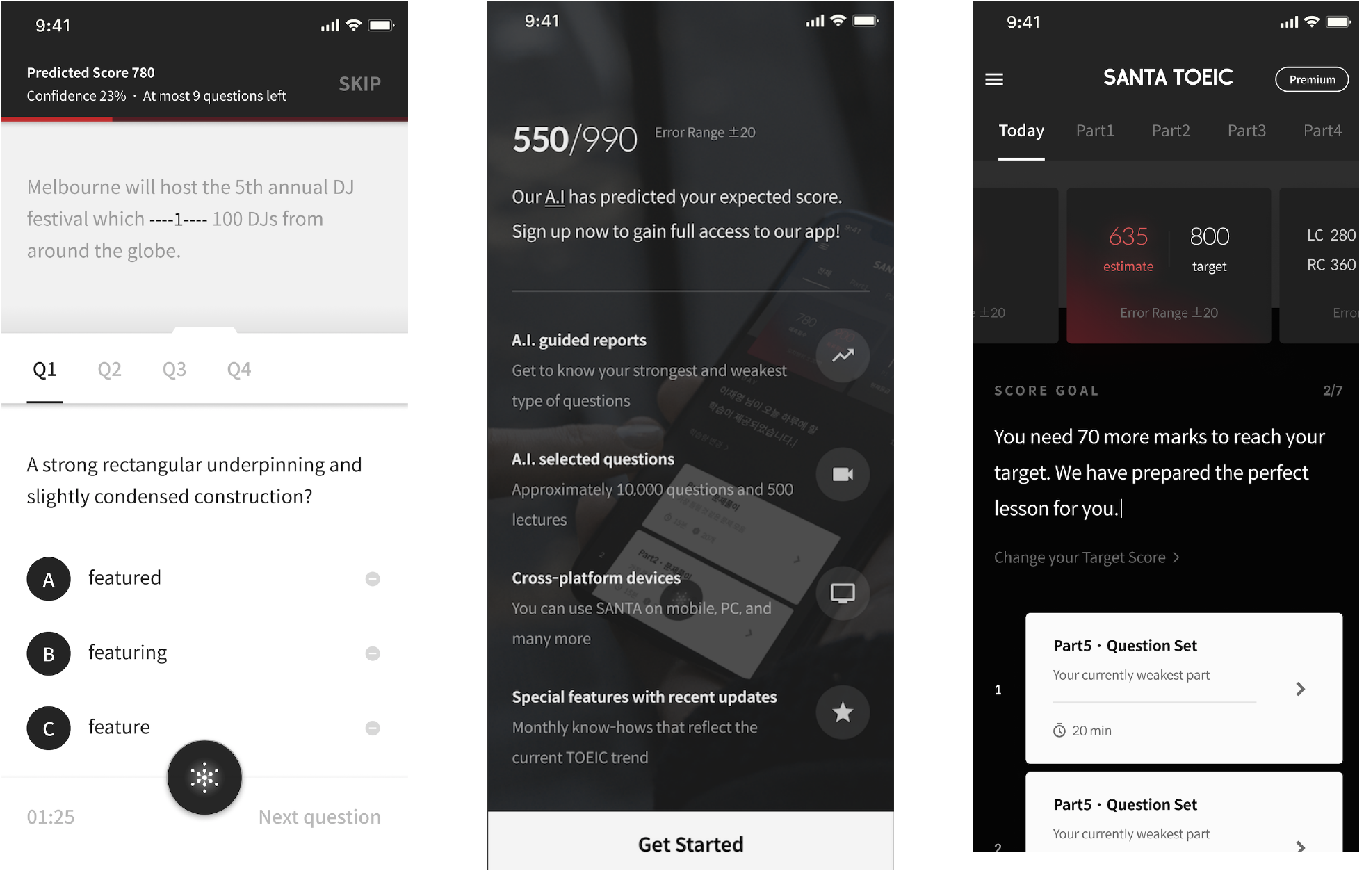}
\end{center}
\caption{Various representations of predicted scores within the application.}
\label{fig:score_model}
\end{figure}

Santa is a multi-platform English ITS with approximately 780K users in South Korea that exclusively focuses on the TOEIC (Test of English for International Communications) standardized examinations. TOEIC is an English proficiency test that consists of two timed sections (listening and reading) each with 100 questions that adds up to a combined total score between 0 to 990. Santa utilizes several AI techniques to optimize the preparation process of the TOEIC examination for students. When the application is first initiated, a preliminary placement test with 7 to 11 problems is given to diagnose the student's current state and predict their expected score in real-time. After the diagnostic test, a user response prediction model is used to dynamically suggest problems which corresponds to the student's current position within the TOEIC ladder. The prediction model is calculated by computing a user's overall correctness rate, eliminating problems that students have answered correctly with high probability and then selecting the best possible content based on expert heuristics. Based on the user's previous data, the predicted scores can be provided in various forms throughout the service, as shown in Figure \ref{fig:score_model}.
Figure \ref{fig:model_flow} shows the flow of score prediction.

\subsection{Performance of Score Prediction Model}
Santa has previously used a CF-based model for score prediction which has recently been replaced with a deep attentive model.
To train the model, we aggregate the real TOEIC scores reported by users of Santa. Santa offered to reward to the users who have reported their score and was able to obtain a total of 2,594 score reports for 6 months. 
The data is then divided into a training set (1,302 users, 1815 labels), validation set (244 users, 260 labels), and a test set (466 users, 519 labels). We use EdNet as pre-training task data and the student sequence data as the label (TOEIC score).
Table \ref{table:MAE} shows the MAE (Mean Absolute Error) of the two models for the test set.

\begin{table}[h]
\centering
\begin{tabular}{ccc}
    & CF & Deep Attentive  model \\
    \hline
MAE & 78.91                   & 49.84           \\
\hline
\end{tabular}
\label{table:MAE}
\caption{MAE of collaborative filtering and attentive  model}
\end{table}

\subsection{A/B test setup}

From February 24th to April 2nd, we conducted an A/B test by randomly administering two different score prediction algorithms to the application users: one based on a collaborative-filtering algorithm and another one based on deep-learning. 50,451 students were allocated to the collaborative filtering algorithm and 17,019 students were provided a deep-learning algorithm. We analyzed each student's response and action (such as time of registration, question response time, purchase rate, etc.) to spot any noteworthy statistics that can validate our experiment. 

\begin{figure*}[t]
\centering
\begin{tikzpicture}
\begin{axis}[
width=15cm,
height=5cm,
xtick=data,
xticklabel style={rotate=40, font=\tiny},
symbolic x coords={02/24, 02/25, 02/26, 02/27, 02/28, 02/29, 03/01, 03/02, 03/03, 03/04, 03/05, 03/06, 03/07, 03/08, 03/09, 03/10, 03/11, 03/12, 03/13, 03/14, 03/15, 03/16, 03/17, 03/18, 03/19, 03/20, 03/21, 03/22, 03/23, 03/24, 03/25, 03/26, 03/27, 03/28, 03/29, 03/30, 03/31, 04/01, 04/02},
]
% 01/26, 01/27, 01/28, 01/29, 01/30, 01/31, 02/01, 02/02, 02/03, 02/04, 02/05, 02/06, 02/07, 02/08, 02/9, 02/10, 02/11, 02/12, 02/13, 02/14
\addplot[color=blue, mark=*]
   coordinates {
(02/24,8.50) (02/25,6.90) (02/26,-3.25) (02/27,6.19) (02/28,3.03) (02/29,0.23) (03/01,-3.32) (03/02,8.32) (03/03,2.46) (03/04,7.98) (03/05,14.21) (03/06,-0.03) (03/07,-0.71) (03/08,9.76) (03/09,1.73) (03/10,9.59) (03/11,7.85) (03/12,-8.96) (03/13,-2.84) (03/14,-8.96) (03/15,-0.26) (03/16,5.78) (03/17,2.51) (03/18,3.45) (03/19,2.46) (03/20,-1.49) (03/21,8.71) (03/22,5.98) (03/23,2.96) (03/24,-2.99) (03/25,0.89) (03/26,2.76) (03/27,4.38) (03/28,-4.63) (03/29,1.58) (03/30,-1.62) (03/31,0.72) (04/01,0.66) (04/02,0.52)
   };
%   (01/26,1.7309756) (01/27,3.4151511) (01/28,2.3872647) (01/29,3.4768657) (01/30,3.8736379) (01/31,-0.3209032) (02/01,1.8056347) (02/02,3.7330553) (02/03,3.2063058) (02/04,-2.8328804) (02/05,0.9921274) (02/06,-0.3740436) (02/07,0.240888) (02/08,-4.6789252) (02/09,0.320351) (02/10,0.3549311) (02/11,0.0570234) (02/12,0.0174123) (02/13,2.3013617) (02/14,3.0211982)
% (02/24,8.50) (02/25,6.90) (02/26,-3.25) (02/27,6.19) (02/28,3.03) (02/29,0.23) (03/01,-3.32) (03/02,8.32) (03/03,2.46) (03/04,7.98) (03/05,14.21) (03/06,-0.03) (03/07,-0.71) (03/08,9.76) (03/09,1.73) (03/10,9.59) (03/11,7.85) (03/12,-8.96) (03/13,-2.84) (03/14,-8.96) (03/15,-0.26) (03/16,5.78) (03/17,2.51) (03/18,3.45) (03/19,2.46) (03/20,-1.49) (03/21,8.71) (03/22,5.98) (03/23,2.96) (03/24,-2.99) (03/25,0.89) (03/26,2.76) (03/27,4.38) (03/28,-4.63) (03/29,1.58) (03/30,-1.62) (03/31,0.72) (04/01,0.66) (04/02,0.52)
\end{axis}
\end{tikzpicture}
\caption{Comparison of the number of questions solved per day between the users of the A/B test.}
\label{fig:changes_u_q}
\end{figure*}
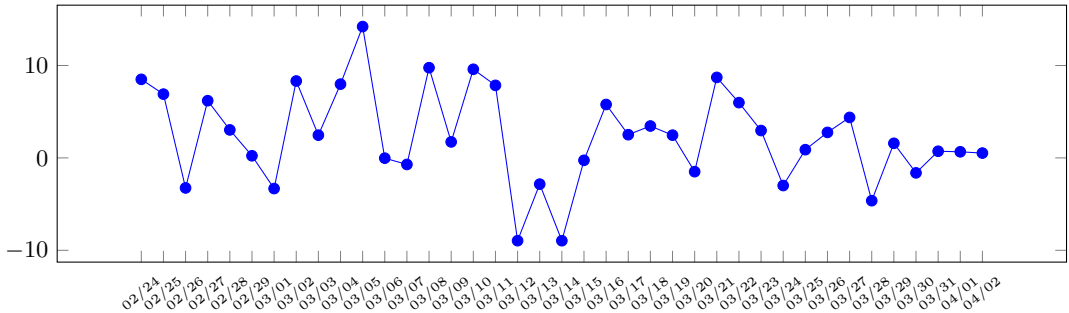
\subsection{Experimental Results}
In this section, we discuss how a high quality of the predicted scores can significantly impact student morale. 

\subsubsection{Student Motivation}
Our first test statistic is the preliminary test completion ratio. The completion rate of the initial placement test is a crucial indicator that could represent a student's motivation, as only students who are willing to learn will try to finish their diagnostic test. For each question a student answers in SANTA, a predicted score that is updated in real-time is projected on the top left corner. This allows for the user to immediately check the quality of the expected score, thus strengthening the trust that the user may have with the application. 
A/B test results show that the deep attentive model has a higher completion rate of 65.90\% than the CF-model with 64.93\%. 

Next, we look at changes in membership rates. A membership rate of an application in a sense signifies greater magnitude of student motivation than the completion rate as it directly indicates the determination of a user who wishes to use the application. Out of a total of 67,470 users that have used Santa during the A/B test period, 44,297 users finished their diagnostic tests and 28,065 users have registered to sign up with the application. The A/B test shows that the deep attentive model has a registration rate of 44.55\% while the CF-based model has 43.13\%.

The average number of questions a user answered after the diagnostic test is also significant proof of a student's educational drive. The A/B test results show that with a deep attentive model a student solved an average of 22.73 questions, while with a CF-based model the user only solved 20.03. Figure \ref{fig:changes_u_q} shows the comparison of the number of questions answered per day between the users of the A/B test. The x-axis represents the date and the y-axis represents the gap between average number of questions answered in a deep attentive model and a CF-based model. If the gap is positive, the former model has on average more questions solved, and vice versa. We can observe that more questions from the deep attentive model were solved mostly throughout the A/B test time period. 

% \begin{tabular}
% \begin{tabularx}{\columnwidth}
\begin{table}[h]
\centering
\begin{tabular}{ccc} 
& CF & Deep Attentive  model \\
\hline
Completion rate (\%) & 64.93 & 65.90 \\
\hline
Registration rate (\%) &43.13 & 44.55  \\
\hline
\# of solved questions & 20.03  & 22.73 \\          
\hline
\end{tabular}
\label{table:motivation}
\caption{Experimental results of student motivation}
\end{table}

\subsubsection{Active Student Engagement}
In this section, we demonstrate active student engagement based on different score prediction models via taking a look at the financial benefits the models bring. Monetary profits are an essential factor in evaluating a service, since it is an important indicator of user engagement as a high level of user engagement directly results in financial success. We measure business impact with 3 metrics : purchase rate, Average Revenue Per User (ARPU), and total profit. In this context, purchase rate is defined as the number of users that decided to purchase full access to the app during the A/B test period. 
The test results show that the purchase rate for the deep attentive model was 2.73\% while the CF-based model had a 2.37\% rate, showing a 15.19\% increase for the deep attentive model. For ARPU, the deep attentive model averaged \$3.23 whilst a CF-based model averaged \$2.83. Total profit during testing period also yielded \$162,933.88 for the former while it only gathered \$142,949.55 for the latter (since the two models had different parameters, these values were normalized based on the ratio of the model parameters). Comparing these 3 metrics, we conclude that the model with higher accuracy in the deep attentive model shows better results as well. 

\begin{table}[h]
\small
\centering
\begin{tabular}{ccc}
& CF & Deep Attentive  model \\
\hline
Conversion rate (\%) & 2.37       & 2.73 \\ 
\hline
ARPU (\$)          & 2.83       & 3.23\\ 
\hline
Total profit (\$)  & 142,949.55 & 162,933.88\\
\hline
\end{tabular}
\label{table:engagement}
\caption{Experimental results of student engagement}
\end{table}

\section{conclusions}
Recent developments in ITS have enabled customized education by suggesting optimal strategies for individual students to approach studying. SANTA has also assisted its users to better prepare for the TOEIC English fluency standardized examinations by utilizing various learning techniques. Recently, SANTA has shifted from a collaborative-filtering model to a deep attentive model that has proved to be an upgrade over the former. To inquire about the benefits of using a fastidious model, this paper conducts various experiments and investigates their results. Analyzing the results of various experiments leads us to believe that deep attentive model entails a higher level of student motivation and engagement. Therefore, we claim that a more accurate model, in this case, the deep attentive model, could induce improved student engagement.

\bibliographystyle{abbrv}
\bibliography{ref}

\begin{thebibliography}{10}

\bibitem{balakrishnan2013predicting}
G.~Balakrishnan and D.~Coetzee.
\newblock Predicting student retention in massive open online courses using
  hidden markov models.
\newblock {\em Electrical Engineering and Computer Sciences University of
  California at Berkeley}, 53:57--58, 2013.

\bibitem{bydvzovska2016course}
H.~Byd{\v{z}}ovsk{\'a}.
\newblock Course enrollment recommender system.
\newblock {\em International Educational Data Mining Society}, 2016.

\bibitem{choi2020assessment}
Y.~Choi, Y.~Lee, J.~Cho, J.~Baek, D.~Shin, S.~Lee, J.~Shin, C.~Bae, B.~Kim, and
  J.~Heo.
\newblock Assessment modeling: Fundamental pre-training tasks for interactive
  educational systems, 2020.

\bibitem{choi2019ednet}
Y.~Choi, Y.~Lee, D.~Shin, J.~Cho, S.~Park, S.~Lee, J.~Baek, C.~Bae, B.~Kim, and
  J.~Heo.
\newblock Ednet: A large-scale hierarchical dataset in education, 2019.

\bibitem{devlin2018bert}
J.~Devlin, M.-W. Chang, K.~Lee, and K.~Toutanova.
\newblock Bert: Pre-training of deep bidirectional transformers for language
  understanding.
\newblock {\em arXiv preprint arXiv:1810.04805}, 2018.

\bibitem{gonzalez2018inferring}
J.~P. Gonz{\'a}lez-Brenes and R.~Edezhath.
\newblock Inferring course enrollment from partial data.
\newblock In {\em International Conference on Artificial Intelligence in
  Education}, pages 429--432. Springer, 2018.

\bibitem{hu2018course}
Q.~Hu and H.~Rangwala.
\newblock Course-specific markovian models for grade prediction.
\newblock In {\em Pacific-Asia Conference on Knowledge Discovery and Data
  Mining}, pages 29--41. Springer, 2018.

\bibitem{hu2019academic}
Q.~Hu and H.~Rangwala.
\newblock Academic performance estimation with attention-based graph
  convolutional networks.
\newblock {\em arXiv preprint arXiv:2001.00632}, 2019.

\bibitem{hu2019reliable}
Q.~Hu and H.~Rangwala.
\newblock Reliable deep grade prediction with uncertainty estimation.
\newblock {\em arXiv preprint arXiv:1902.10213}, 2019.

\bibitem{iqbal2017machine}
Z.~Iqbal, J.~Qadir, A.~N. Mian, and F.~Kamiran.
\newblock Machine learning based student grade prediction: A case study.
\newblock {\em arXiv preprint arXiv:1708.08744}, 2017.

\bibitem{jo2018time}
Y.~Jo, K.~Maki, and G.~Tomar.
\newblock Time series analysis of clickstream logs from online courses.
\newblock {\em arXiv preprint arXiv:1809.04177}, 2018.

\bibitem{labarthe2016does}
H.~Labarthe, F.~Bouchet, R.~Bachelet, and K.~Yacef.
\newblock Does a peer recommender foster students' engagement in moocs?.
\newblock {\em International Educational Data Mining Society}, 2016.

\bibitem{lee2016machine}
K.~Lee, J.~Chung, Y.~Cha, and C.~Suh.
\newblock Machine learning approaches for learning analytics: Collaborative
  filtering or regression with experts?
\newblock In {\em NIPS Workshop, Dec}, pages 1--11, 2016.

\bibitem{li2016understanding}
Q.~Li and R.~Baker.
\newblock Understanding engagement in moocs.
\newblock In {\em EDM}, pages 605--606, 2016.

\bibitem{lui2017exploring}
J.~Lui and H.~Li.
\newblock Exploring the relationship between student pre-knowledge and
  engagement in moocs using polytomous irt.
\newblock In {\em Proceedings of the 10th International Conference on
  Educational Data Mining}, pages 410--411. ERIC, 2017.

\bibitem{morsy2017cumulative}
S.~Morsy and G.~Karypis.
\newblock Cumulative knowledge-based regression models for next-term grade
  prediction.
\newblock In {\em Proceedings of the 2017 SIAM International Conference on Data
  Mining}, pages 552--560. SIAM, 2017.

\bibitem{morsy2019sparse}
S.~Morsy and G.~Karypis.
\newblock Sparse neural attentive knowledge-based models for grade prediction.
\newblock {\em arXiv preprint arXiv:1904.11858}, 2019.

\bibitem{mulqueeny2015improving}
K.~Mulqueeny, L.~A. Mingle, V.~Kostyuk, R.~S. Baker, and J.~Ocumpaugh.
\newblock Improving engagement in an e-learning environment.
\newblock In {\em International Conference on Artificial Intelligence in
  Education}, pages 730--733. Springer, 2015.

\bibitem{naik2019analyzing}
V.~Naik and V.~Kamat.
\newblock Analyzing engagement in an on-line session.
\newblock In {\em International Conference on Artificial Intelligence in
  Education}, pages 359--364. Springer, 2019.

\bibitem{okur2017behavioral}
E.~Okur, N.~Alyuz, S.~Aslan, U.~Genc, C.~Tanriover, and A.~A. Esme.
\newblock Behavioral engagement detection of students in the wild.
\newblock In {\em International Conference on Artificial Intelligence in
  Education}, pages 250--261. Springer, 2017.

\bibitem{patil2017effective}
A.~P. Patil, K.~Ganesan, and A.~Kanavalli.
\newblock Effective deep learning model to predict student grade point
  averages.
\newblock In {\em 2017 IEEE International Conference on Computational
  Intelligence and Computing Research (ICCIC)}, pages 1--6. IEEE, 2017.

\bibitem{polyzou2016grade}
A.~Polyzou and G.~Karypis.
\newblock Grade prediction with course and student specific models.
\newblock In {\em Pacific-Asia Conference on Knowledge Discovery and Data
  Mining}, pages 89--101. Springer, 2016.

\bibitem{rechkoski2018evaluation}
L.~Rechkoski, V.~V. Ajanovski, and M.~Mihova.
\newblock Evaluation of grade prediction using model-based collaborative
  filtering methods.
\newblock In {\em 2018 IEEE Global Engineering Education Conference (EDUCON)},
  pages 1096--1103. IEEE, 2018.

\bibitem{ren2019grade}
Z.~Ren, X.~Ning, A.~Lan, and H.~Rangwala.
\newblock Grade prediction based on cumulative knowledge and co-taken courses.
\newblock In {\em Proceedings of the 12th International Conference on
  Educational Data Mining (EDM)}. ERIC, 2019.

\bibitem{ren2018ale}
Z.~Ren, X.~Ning, and H.~Rangwala.
\newblock Ale: Additive latent effect models for grade prediction.
\newblock In {\em Proceedings of the 2018 SIAM International Conference on Data
  Mining}, pages 477--485. SIAM, 2018.

\bibitem{slater2016semantic}
S.~Slater, R.~Baker, J.~Ocumpaugh, P.~Inventado, P.~Scupelli, and N.~Heffernan.
\newblock Semantic features of math problems: Relationships to student learning
  and engagement.
\newblock {\em International Educational Data Mining Society}, 2016.

\bibitem{slim2018predicting}
A.~Slim, D.~Hush, T.~Ojah, and T.~Babbitt.
\newblock Predicting student enrollment based on student and college
  characteristics.
\newblock {\em International Educational Data Mining Society}, 2018.

\bibitem{tadayon2019predicting}
M.~Tadayon and G.~Pottie.
\newblock Predicting student performance in an educational game using a hidden
  markov model.
\newblock {\em arXiv preprint arXiv:1904.11857}, 2019.

\bibitem{vaswani_2017}
A.~Vaswani, N.~Shazeer, N.~Parmar, J.~Uszkoreit, L.~Jones, A.~N. Gomez,
  {\L}.~Kaiser, and I.~Polosukhin.
\newblock Attention is all you need.
\newblock In {\em Advances in neural information processing systems}, pages
  5998--6008, 2017.

\bibitem{warner2015high}
J.~Warner, J.~Doorenbos, B.~Miller, and P.~J. Guo.
\newblock How high school, college, and online students differentially engage
  with an interactive digital textbook.
\newblock In {\em EDM}, pages 528--531, 2015.

\bibitem{yin2019quesnet}
Y.~Yin, Q.~Liu, Z.~Huang, E.~Chen, W.~Tong, S.~Wang, and Y.~Su.
\newblock Quesnet: A unified representation for heterogeneous test questions.
\newblock {\em arXiv preprint arXiv:1905.10949}, 2019.

\end{thebibliography}

\end{document}